\documentstyle[aps,multicol,epsf]{revtex}
\input epsf
\draft
\title{Suppression of Magnetic State Decoherence Using Ultrafast Optical Pulses}
\author{C. Search and P. R. Berman} 
\address{Physics Department, University of Michigan, Ann Arbor, MI \ 48109-1120}
\date{\today }
\begin{document}
\maketitle
\pacs{32.80.Qk, 34.50.Rk, 34.20.Cf}
\begin{abstract}
It is shown that the magnetic state decoherence produced by collisions in a
thermal vapor can be suppressed by the application of a train of ultrafast
optical pulses.
\end{abstract}
\begin{multicols}{2}
\narrowtext

In a beautiful experiment, Itano {\it et al.} demonstrated the Quantum Zeno 
{\em effect }\cite{itano}. A radio frequency pi pulse having a duration on
the order of 250 ms was applied to a ground state hyperfine transition. At
the same time, a series of radiation{\em \ }pulses was used to drive a
strongly coupled ground to excited state {\em uv} transition. The {\em rf}
and strong transitions shared the same ground state level. Itano {\it et al. 
}showed that excitation of the {\em rf} transition could be suppressed by
the {\em uv} pulses. They interpreted the result in terms of collapse of the
wave function - spontaneous emission from the excited state during the {\em %
uv }pulses is a signature that the {\em uv} pulse projected the atom into
its ground state; the lack of such spontaneous emission implies projection
into the final state of the {\em rf} transition. This paper triggered a
great deal of discussion, especially with regards to the {\em interpretation}
of the results \cite{zenorev}.

A necessary condition for a quantum Zeno effect is a perturbation of a state
amplitude on a time scale that {\em is short compared with the correlation
time of the process inducing the transition}. In the experiment of Itano 
{\it et al., }this time scale is simply the duration of the pi pulse, 256
ms. On the other hand, if one wished to inhibit particle decay or
spontaneous emission \cite{sundar}, it would be necessary to apply
perturbations on a time scale that is short compared with the correlation
time of the vacuum, an all but impossible task. In this paper,we consider
the inhibition of collisional, magnetic state decoherence, by the
application of a train of ultrafast, optical pulses. This correlation time
of the collisional perturbations resulting in magnetic state decoherence is
of order of the duration of a collision and is intermediate between that for
the coherent pi pulse applied by Itano {\it et al. }and the almost
instantaneous, quantum jump-like process produced by the vacuum field. It
should be noted that related schemes have been proposed for inhibiting
decoherence in systems involving quantum computation \cite{quant}, but the
spirit of these proposals differs markedly from the one presented herein.

The rapid perturbations of the system are a necessary, but not sufficient,
condition for a mechanism to qualify as a Quantum Zeno effect. The
perturbations must involve some ''measurement'' on the system for the
''Quantum Zeno'' label to apply. The suppression of magnetic state coherence
discussed in this paper is not a Quantum Zeno effect in this sense. We will
return to this point below.

\begin{figure}[tb!]
\centering
\begin{minipage}{8.0cm}
\epsfxsize= 8 cm \epsfysize= 8.8 cm \epsfbox{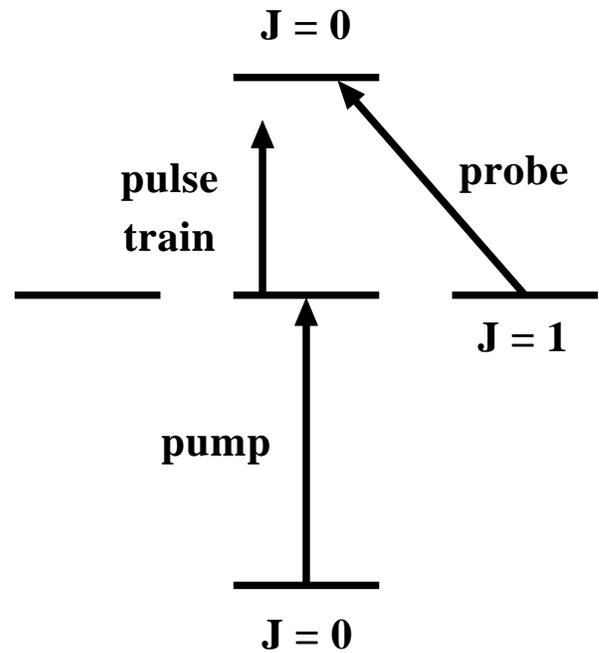}
\end{minipage}
\caption{Energy level diagram. The collisional interaction couples the
magnetic sublevels in the $J=1$ state.}
\label{znfig1}
\end{figure}

We envision an experiment in which ''active atoms'' in a thermal vapor
undergo collisions with a bath of foreign gas perturbers. A possible level
scheme for the active atoms is depicted in Fig. \ref{znfig1}. At some
initial time, an ultrashort pulse excites an atom from its ground state,
having angular momentum $J=0,$ to the $m=0$ sublevel of an excited state
having $J=1$. The duration of the excitation pulse $\tau _{p}$ is much
shorter than the {\em duration} of a collision $\tau _{c}$ ($\tau _{c}$ is
typically of order $1$ ps). As a result of elastic collisions with the
ground state perturbers, population in the $J=1$ sublevels equilibrate at a
rate $\Gamma _{col}$ that is typically of order $10^{7}-10^{8}$ s$^{-1}$ per
Torr of perturber pressure. The transfer to the $m=1$ substate is probed by
a circularly polarized pulse acting on the $J=1,m=0\rightarrow J=0$ excited
state transition, applied at a time $\Gamma _{col}^{-1}$ following the
initial excitation pulse. For the sake of definiteness, we assume that the
perturber pressure is such that equilibration occurs in a time $\Gamma
_{col}^{-1}=0.1-1.0$ ns. The question that we address in this paper is ''How
can one inhibit this magnetic state decoherence by subjecting the active
atoms to additional external radiation fields?''

As was mentioned above, the key to any Zeno-type effect is to disrupt the
coherent evolution a system from its initial to final state. In our case,
the coherent evolution from the initial $m=0$ states to the final $m=\pm 1$
states is driven by the collisional interaction. Thus it is necessary to
disturb the system on a time scale that is short compared with the collision
duration $\tau _{c}$. To do this, we apply a continuous train of ultrashort
pulses that couple the $m=0$ level to the excited state having $J=0$ shown
in Fig. \ref{znfig1}. The pulses are assumed to have duration $\tau _{p}\ll
\tau _{c}$ and are assumed to be off-resonance; that is the atom-field
detuning is large compared with $\tau _{p}^{-1}$. As such, each pulse simply
produces an $ac$ Stark shift of the $m=0$ sublevel of the $J=1$ state,
resulting in a {\em phase shift }of this state amplitude. As a consequence,
the external pulses break the collision-induced, coherent evolution of the
atom from its initial $m=0$ state to the $m=\pm 1$ states. If the pulse
strengths are chosen such that the phase shift is a random number, modulo $%
2\pi ,$ and if many pulses occur during the collision duration $\tau _{c}$,
then the atom will be frozen in its initial state. It is interesting to note
that collisions, which are normally viewed as a decohering process, must be
viewed as a {\em coherent} driving mechanism on the time scales considered
in this work.

To obtain a qualitative understanding of this effect, it is sufficient to
consider a model, two-level system, consisting of an initial state $\left|
0\right\rangle $ (corresponding to the $J=1,$ $m=0$ state) and a final state 
$\left| 1\right\rangle $ (corresponding to the $J=1$, $m=1$ state, for
example). The Hamiltonian for this two-state system is taken as 
\begin{equation}
H=V_{c}(t)\left( \left| 0\right\rangle \left\langle 1\right| +\left|
1\right\rangle \left\langle 0\right| \right) +\hbar \sum_{i}\Delta
_{s}(t_{i})\tau _{p}\delta (t-t_{i})\left| 0\right\rangle \left\langle
0\right| ,  \label{1}
\end{equation}
where $V_{c}(t)$ is a collisional perturbation that couples the two,
degenerate states, and $\Delta _{s}(t_{i})$ is the ac Stark shift of state $%
\left| 0\right\rangle $ produced by the external pulse occurring at $t=t_{i}$%
. For simplicity, we take $V_{c}(t)$ to be a square pulse, $V_{c}(t,b)=\hbar
\beta (b),$ for\ 0$\leq t\leq \tau _{c}$. The quantity $b$ is the impact
parameter of the collision. Without loss of generality, we can take the
collision to start at $t=0$. The collision duration $\tau _{c}$ can be
written in terms of the impact parameter $b$ characterizing the collision
and the relative active atom-perturber speed $u$ as $\tau _{c}(b)=b/u.$
Moreover, to simulate a van der Waals interaction, we set $\beta
(b)=(C/b_{0}^{6})\left( b_{0}/b\right) ^{6},$ where $C$ and $b_{0}$ are
constants chosen such that $2C/(b_{0}^{5}u)=1.$ The quantity $b_{0}$ is an
effective Weisskopf radius for this problem. An average over $b$ will be
taken to calculate the transition rate.

The external pulse train is modeled in two ways. In model A, the pulses
occur at random times with some average separation $T$ between the pulses.
In model B, the pulses are evenly spaced with separation $T.$ In both
models, the pulse areas $\Delta _{s}(t_{i})\tau _{p}$ are taken to be random
numbers between 0 and 2$\pi .$ A quantity of importance is the average
number of pulses, $n_{0}=\tau _{c}(b_{0})/T=b_{0}/(uT),$ for a collision
having impact parameter $b_{0}$.

\subsection{Randomly-spaced pulses}

The randomly spaced, radiative pulses act on this two-level system in a
manner analogous to the way collisions modify atomic electronic-state
coherence. In other word, the pulses do not affect the state populations,
but {\em do} modify the coherence between the levels. The pulses can be
treated in an impact approximation, such that {\em during} a collision, the
time rate of change of density matrix elements resulting from the pulses is $%
\dot{\rho}_{00}=\dot{\rho}_{11}=0$ and 
\begin{equation}
\dot{\rho}_{10}/\rho _{10}=\dot{\rho}_{01}/\rho _{01}=-\Gamma \left\langle
1-e^{-i\Delta _{s}(t_{i})\tau _{p}}\right\rangle =-\Gamma ,
\end{equation}
where $\Gamma =T^{-1}$ is the average pulse rate and we have used the fact
that the pulse area is a random number between 0 and 2$\pi $. Taking into
account the collisional coupling $V_{c}(t,b)$ between the levels, one
obtains evolution equations for components of the Bloch vector $w=\rho
_{11}-\rho _{00}=2\rho _{11}-1$, $v=i(\rho _{10}-\rho _{01})$ as 
\begin{equation}
dw/dx=U(y)v;\text{ \ \ \ }dv/dx=-U(y)w-n(y)v,  \label{2}
\end{equation}
where $x=t/\tau _{c}(b)$ is a dimensionless time, $y=b/b_{0}$ is a relative
impact parameter$,$ and $U(y)=y^{-5}$ and $n(y)=n_{0}y$ are dimensionless
frequencies. These equations are solved subject to the initial condition $%
w(0)=-1;$ $v(0)=0$, to obtain the value $\rho
_{11}(x=1,y,n_{0})=[w(x=1,y)+1]/2.$ The relative transition rate $S$ is
given by 
\begin{equation}
S(n_{0})=2\pi Nub_{0}^{2}\int_{0}^{\infty }y\,dy\,\rho _{11}(x=1,y,n_{0})/2,
\label{3}
\end{equation}
where $N$ is the perturber density. A coefficient, $R(n_{0}),$ which
measures the suppression of decoherence, can be defined as 
\begin{equation}
R(n_{0})=\int_{0}^{\infty }y\,dy\,\rho _{11}(x=1,y,n_{0})/\int_{0}^{\infty
}y\,dy\,\rho _{11}(x=1,y,0)  \label{4}
\end{equation}

Solving Eqs. (\ref{2}), one finds 
\begin{mathletters}
\label{5}
\begin{eqnarray}
\rho _{11}(x &=&1,y,n_{0})=\left[ 1-\frac{r_{1}}{r_{2}-r_{1}}\left(
e^{-r_{1}}-\frac{r_{1}}{r_{2}}e^{-r_{2}}\right) \right] /2;  \label{5a} \\
r_{1,2} &=&\left( -n_{0}y\pm \sqrt{(n_{0}y)^{2}-4y^{-10}}\right) /2.
\label{5b}
\end{eqnarray}
It is now an easy matter to numerically integrate Eqs. (\ref{4}) to obtain $%
R(n_{0})$. Before presenting the numerical results, we can look at some
limiting cases which provide insight into the physical origin of the
suppression of decoherence.

A plot of $\rho _{11}(x=1,y,n_{0})$ as a function of $y=b/b_{0}$ is shown in
Fig. \ref{fig2} for several values of $n_{0}.$ With decreasing $y$, $\rho
_{11}$ increases monotonically to some maximum value $\rho _{11}(y_{m})$ and
then begins to oscillate about $\rho _{11}=1/2$ with increasing amplitude.
One concludes from such plots that {\em two} effects contribute to the
suppression of coherence. The first effect, important for large $n_{0}$, is
a reduction in the value of $y_{m}$. The second effect, important for $n_{0}$
of order unity, is a decrease in the value of $\rho _{11}(y_{m})$. Let us
examine these two effects separately.

\begin{figure}[tb!]
\centering
\begin{minipage}{8.0cm}
\epsfxsize= 8 cm \epsfysize= 5.75 cm \epsfbox{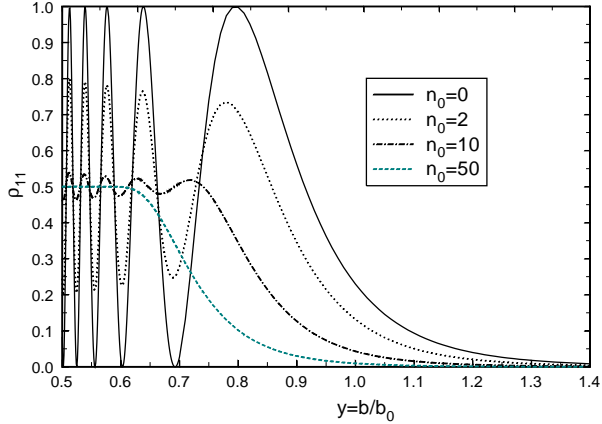}
\end{minipage}
\caption{Graph of $\protect\rho _{11}$ as a function of $y=b/b_{0}$ for
several values of $n_{0}$. For values $0\leq y\leq 0.45$ not shown on the
graph, $\protect\rho _{11}$ oscillates about an average value of 1/2. For $%
n_{0}\neq 0,$ the oscillation amplitude increases with decreasing $y.$}
\label{fig2}
\end{figure}

For very large $n_{0},$ $n_{0}^{5/66}\gg 1,$ one can approximate $\rho _{11}$
over the range of $y$ contributing significantly to the integral (\ref{3})
as $\rho _{11}(x=1,y,n_{0})\sim \left( 1-e^{-y^{-11}/n_{0}}\right) /2.$ By
evaluating the integrals in (\ref{4}), one finds a suppression of
decoherence ratio given by 
\end{mathletters}
\begin{equation}
R(n_{0})=0.95/n_{0}^{2/11}.  \label{6}
\end{equation}
The $n_{0}^{-2/11}$ dependence is a general result for a collisional
interaction that varies as the interatomic separation to the minus 6th
power. It can be understood rather easily. The pulses break up the collision
into $n_{0}y$ segments, each having a (dimensionless) time duration $%
x_{b}=1/(n_{0}y)$. Each segment provides a perturbative contribution to $%
\rho _{11}$ of order $y^{-10}(n_{0}y)^{-2}$, provided $y<y_{w}$, where $%
y_{w} $ is to be determined below. The total population from the entire
collision interval varies as $\rho _{11}\sim
y^{-10}(n_{0}y)^{-2}n_{0}y=y^{-11}/n_{0}$. Of course, $\rho _{11}$ cannot
exceed unity. One can define an effective relative Weisskopf radius, $y_{w},$
as one for which $\rho _{11}=1$, namely $y_{w}=b_{w}/b_{0}=n_{0}^{-1/11}.$
The total transition rate varies as $y_{w}^{2}\sim n_{0}^{-2/11}$, in
agreement with (\ref{6}). As $n_{0}\sim \infty $, the atom is frozen in its
initial state.

\begin{figure}[tb!]
\centering
\begin{minipage}{8.0cm}
\epsfxsize= 8 cm \epsfysize= 5.5 cm \epsfbox{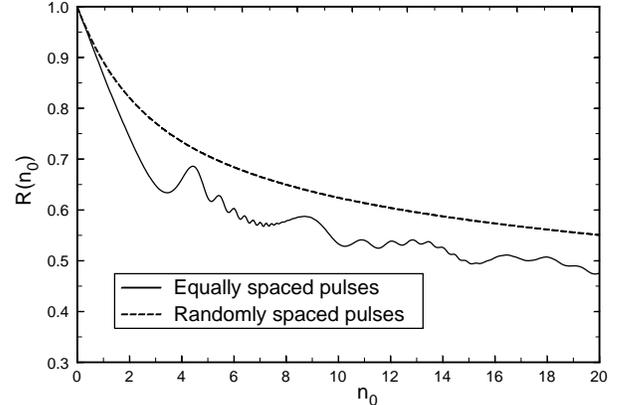}
\end{minipage}
\caption{Graph of the suppression of decoherence ratio $R$ as a function of $%
n_{0}$ for randomly and uniformly spaced pulses.}
\label{fig3}
\end{figure}

For values of $n_{0}$ of order unity, the dominant cause of the suppression
of decoherence is a decrease in the value of $\rho _{11}(y_{m})$, rather
than the relatively small decrease in $y_{m}$ from its value when $n_{0}=0$.
For values $n_{0}\leq 3,$ approximately 45\% of the contribution to the
transition rate $S(n_{0})$ originates from $y>y_{m},$ and, for these values
of $n_{0}$, $y_{m}\sim \pi ^{-1/5}$ and $\rho _{11}(y_{m})\sim
(1+e^{-n_{0}/2\pi ^{1/5}})/2$. This allows us to estimate the suppression of
decoherence ratio as $R(n_{0})=[0.55+.45(1+e^{-n_{0}/2\pi ^{1/5}})/2],$ such
that $R(1)=0.93$, $R(2)=0.88$, $R(3)=0.84.$ These values are approximately
70\% of the corresponding numerical results, indicating that the decrease in 
$\rho _{11}(y_{m})$ accounts for approximately 70\% of the suppression at
low $n_{0}$, with the remaining 30\% coming from a decrease in $y_{m}$. The
first few collisions are relatively efficient in suppressing decoherence.
With increasing $n_{0}$, the suppression process slows, varying as $%
n_{0}^{-2/11}$. In Fig. \ref{fig3}, the suppression of decoherence ratio $%
R(n_{0})$, obtained by a numerical solution of Eq. (\ref{4}), is plotted as
a function of $n_{0}$.

\subsection{Uniformly Spaced Pulses}

We consider now the case of equally spaced pulses, having effective pulse
areas that are randomly chosen, modulo 2$\pi $. The time between pulses is $T
$, and $n_{0}=\tau _{c}(b_{0})/T$. For a relative impact parameter $y=b/b_{0}
$, with $m\leq $ $n(y)=n_{0}y$ $\leq m+1$, where $m$ is a positive integer
or zero, exactly $m$ or $m+1$ pulses occur. The effect of the pulses is
calculated easily using the Bloch vector. At $x=0$, $w=-1$ and $v=0$. The
Bloch vector then undergoes free evolution at frequency $U(y)=$ $y^{-5}$ up
until the (dimensionless) time of the first pulse, $x_{s}=t_{s}/\tau _{c}(b)$%
. The pulse randomizes the phase of the Bloch vector, so that the average
Bloch vector following the pulse is projected onto the $w$ axis. From $%
x=x_{s}$ to $x_{s}$ $+T/\tau _{c}(b)=x_{s}$ $+1/n(y)$, the Bloch vector
again precesses freely and acquires a phase $UT=y^{-5}/n(y)=y^{-6}/n_{0}$,
at which time the next pulse projects it back onto the $w$ axis. Taking into
account the periods of free precession and projection, and averaging over
the time $x_{s}$ at which the {\em first} pulse occurs, one finds 
\begin{eqnarray}
w(y) &=&[1-n(y)]\cos [y^{-5}]  \nonumber \\
&&+n(y)\int_{0}^{1}dx_{s}\,\cos [y^{-5}x_{s}]\cos [y^{-5}(1-x_{s})]; 
\nonumber \\
0 &\leq &y\leq 1/n_{0},  \nonumber \\
w(y) &=&[m+1-n(y)][(m+1)/n(y)-1]^{-1}  \nonumber \\
&&\times \int_{1-m/n(y)}^{1/n(y)}dx_{s}\,\cos [y^{-5}x_{s}]\cos
^{m-1}[y^{-6}/n_{0}]  \nonumber \\
&&\times \cos [y^{-5}\{1-x_{s}-(m-1)/n(y)\}]  \nonumber \\
&&+[n(y)-m]\left[ 1-m/n(y)\right] ^{-1}  \nonumber \\
&&\times \int_{0}^{1-m/n(y)}dx_{s}\,\cos [y^{-5}x_{s}]\cos ^{m}[y^{-6}/n_{0}]
\nonumber \\
&&\times \cos [y^{-5}\{1-x_{s}-m/n(y)\}];  \nonumber \\
\text{\ }m/n_{0}\text{\ } &\leq &y\leq (m+1)/n_{0}\text{\ }\ \text{\ \ \ for 
}m\geq 1.  \label{7}
\end{eqnarray}

In the limit that $n_{0}\gg 1$, for all impact parameters that contribute
significantly to the transition rate, approximately $n(y)$ pulses occur at
relative impact parameter $y$, implying that $w(y)\sim \cos
^{n(y)}[y^{-5}/n(y)]$ and 
\begin{eqnarray}
R(n_{0}) &=&\frac{\left\langle 1-\cos ^{n_{0}y}[y^{-6}/n_{0}]\right\rangle }{%
\left\langle 1-\cos [y^{-5}]\right\rangle }  \label{8} \\
&\sim &\frac{\left\langle 1-[1-y^{-12}/2n_{0}^{2}]^{n_{0}y}\right\rangle }{%
\left\langle 1-\cos [y^{-5}]\right\rangle }  \nonumber \\
&\sim &\frac{\left\langle 1-e^{-y^{-11}/2n_{0}}\right\rangle }{\left\langle
1-\cos [y^{-5}]\right\rangle }=\frac{0.84}{n_{0}^{2/11}},  \nonumber
\end{eqnarray}
which is the same functional dependence found for the randomly spaced
pulses. Note that the form \{$1-\cos ^{n(y)}[y^{-15}/n(y)]\}$ is identical
to that found in theories of the Zeno effect \cite{itano}.

The suppression of decoherence ratio $R(n_{0})$, obtained from Eqs. (\ref{4}%
) and (\ref{7}) [using $\rho _{11}=(1+w)/2]$, is plotted in Fig. 3. The fact
that it lies below that for randomly spaced pulses is connected with the
difference in the average collisional phase shift acquired between radiation
pulses for the two models. The oscillations in $R(n_{0})$ appear to be an
artifact of our square pulse collision model. In the {\em absence} of the
pulses, the first maximum in the transition cross section occurs for $%
y_{\max }=(\pi )^{-1/5}$, corresponding to a $\pi $ collision pulse. With
increasing $n_{0}$, the pulses divide the collision duration into
approximately $n(y)$ equal intervals. If these pulse intervals are odd or
even multiples of $\pi ,$ one can enhance or suppress the contribution to
the transition rate at specific impact parameters. Numerical calculations
carried out for a smooth interatomic potential do not exhibit these
oscillations.

\subsection{Discussion}

Although the collisional interaction has been modeled as a square pulse, the
qualitative nature of the results is unchanged for a more realistic
collisional interaction, including level shifts. In fact, for a smooth
interatomic potential that allows for an increased number of radiation
pulses over the duration of the collisional interaction, the suppression is
slightly enhanced from the square pulse values. Although the pulses are
assumed to drive only the $J=1,m=0\rightarrow $ $J=0,$ excited state
transition, it is necessary only that the incident pulses produce different
phase shifts on the $J=1,m=0$ and $J=1,m=1$ state amplitudes.

To observe the suppression of decoherence, one could use Yb as the active
atom and Xe perturbers. The Weisskopf radius for magnetic decoherence is
about 1.0 nm \cite{legouet}$,$ yielding a decoherence rate of $\simeq
10^{10} $ s$^{-1}$ at 500 Torr of Xe pressure at 300$^{\circ }C$, and a
collision duration $\tau _{c}(b_{0})$ $\simeq 2.5$ ps. Thus, by choosing a
pulse train having pulses of duration $\tau _{p}=$100 fs, separated by 0.5
ps, it is possible to have 5 pulses per collision. If an experiment is
carried out with an overall time of 100 ps (time from initial excitation to
probing of the final state), one needs a train of about 200 pulses. To
achieve a phase shift $\Delta _{s}\tau _{p}$ of order $2\pi $ and maintain
adiabaticity, one can take the detuning $\delta =3\times $10$^{13}$ s$^{-1}$
and the Rabi frequency $\Omega \simeq 1\times $10$^{14}$ s$^{-1}$ on the $%
J=1,m=0\rightarrow $ $J=0,$ excited state transition \cite{phase}. The
corresponding, power density is $\simeq 1.5\times 10^{11}$ W/cm$^{2}$, and
the power per pulse is $\simeq 150$ $\mu $J (assuming a $1$ mm$^{2}$ focal
spot size). This is a rather modest power requirement. With 5
pulses/collision duration, one can expect a relative suppression of magnetic
state decoherence of order 40\%.

Finally, we should like to comment on whether or not the effect described in
this work constitutes a Quantum Zeno effect. Normally, the Quantum Zeno
effect is presented as a projection of a quantum system onto a given state
as a result of a measurement on the system. In the experiment of Itano et
al., this ''measurement'' is reflected by the presence or absence of
spontaneously emitted radiation during each {\em uv }''measurement'' pulse.
The measurement pulse must be sufficiently long to produce a high likelihood
of spontaneous emission whenever the atom is ''projected'' into the initial
state by the pulse. Following each measurement pulse, the off-diagonal
density matrix element for the two states of the {\em rf} transition goes to
zero. In our experiment involving {\em off-resonant} pulses, the number of
Rayleigh photons scattered from the $J=0$ level during each applied pulse is
much less than unity. As such, there is no Quantum Zeno effect, even if
suppression of magnetic state decoherence occurs. {\em On average}, each
pulse having random area destroys the coherence between the $J=1,m=0$ and $%
J=1,m=\pm 1$ state amplitudes, but does not kill this coherence for a {\em %
single} atom. With an increasing number of radiation pulses, $n_{0}$,
however, both the average value and the standard deviation of the transition
probability tends to zero as $n_{0}^{-1}$ for each atom in the ensemble.

The experiment of Itano {\it et al}. could be modified to allow for a
comparison with the theory presented herein, and to observe the transition
into the Quantum Zeno regime. If the pulses that drive the strong transition
are replaced by a sequence of off-resonant pulses, each pulse having a
duration $\tau _{p}$ much less than the time, $T_{\pi }$, required for the
pi pulse to drive the weak transition, and each pulse having an effective
area, $\Delta _{s}\tau _{p}=(\Omega ^{2}/4\delta )\tau _{p}$, that is random
in the domain [0,2$\pi ],$ then the pulses will suppress the excitation of
the weak transition (it is assumed that $\Omega /\delta \ll 1)$. If the
upper state decay rate is $\gamma $, then the average number of Rayleigh
photons scattered during each pulse is $n=\left( \Omega /4\delta \right)
^{2}\gamma \tau _{p}.$ For $n<1$, there is suppression of the transition
rate as in our case, while, for $n\gtrsim 1$, there is suppression {\em and}
a Quantum Zeno effect. There is no average over impact parameter, since
exactly $[T_{\pi }/T$] or ([$T_{\pi }/T$]+1) pulses in each interval between
the pulses, where [$x$] indicates the integer part of $x$.

\subsection{Acknowledgments}

PRB is pleased to acknowledge helpful discussions with R. Merlin, A. Rojo
and J. Thomas. This research is supported by the National Science Foundation
under grant PHY-9800981 and by the U. S. Army Research Office under grants
DAAG55-97-0113 and DAAH04-96-0160.

\end{multicols}

\begin{references}
\bibitem{phase}  In the adiabatic limit, the phase is equal to $(1/2)\int \{%
\sqrt{[\Omega (t)]^{2}+\delta ^{2}}-\delta \}\,dt$.
\end{references}
\end{document}